\documentclass[aps,pre,twoside,floatfix,superscriptaddress,showpacs,twocolumn]{revtex4}
\usepackage{amsmath}
\usepackage{amsfonts}
\usepackage{bbm}
\usepackage{graphicx}
\usepackage{overpic}
\usepackage{psfrag}

\begin{document}

\title{Capillary interactions in Pickering emulsions}

\author{J.~Guzowski}
\email{jguzowski@ichf.edu.pl}
\affiliation{Institute of Physical Chemistry, Polish Academy of Sciences, ul.\ Kasprzaka 44/52 01--224, Warsaw, Poland}
\author{M.~Tasinkevych}
\affiliation{Max-Planck-Institut f\"ur Intelligente Systeme, Heisenbergstr.\ 3, 70569 Stuttgart, Germany}
\affiliation{Institut f\"{u}r Theoretische und Angewandte Physik, Universit\"{a}t Stuttgart, Pfaffenwaldring 57, D-70569 Stuttgart, Germany}
\author{S.~Dietrich}
\affiliation{Max-Planck-Institut f\"ur Intelligente Systeme, Heisenbergstr.\ 3, 70569 Stuttgart, Germany}
\affiliation{Institut f\"{u}r Theoretische und Angewandte Physik, Universit\"{a}t Stuttgart, Pfaffenwaldring 57, D-70569 Stuttgart, Germany}

\date{\today}

\begin{abstract}
The effective capillary interaction potentials for small colloidal particles trapped at the surface of liquid droplets are calculated analytically. Pair potentials between capillary monopoles and dipoles, corresponding to particles floating on a droplet with a fixed center of mass and subjected to external forces and torques, respectively, exhibit a repulsion at large angular separations and an attraction at smaller separations, with the latter resembling the typical behavior for flat interfaces. This change of character is not observed for quadrupoles, corresponding to free particles on a mechanically isolated droplet. The analytical results for quadrupoles are compared with the numerical minimization of the surface free energy of the droplet in the presence of ellipsoidal particles. 
\end{abstract}

\pacs{68.03.Cd, 83.80.Iz, 83.80.Hj}

\maketitle
 

\section{Introduction}

If colloidal particles get trapped at fluid-fluid interfaces they interact effectively via deformations of the interface. These so-called capillary interactions can easily be tuned by changing external fields and they depend sensitively on the shape of the particles. This makes them good candidates for designing self-assembling systems and provides a convenient experimental playground for studying basic issues of statistical mechanics in two dimensions in the presence of long-ranged interactions~\cite{Zahn2000,Dominguez2010}. On the other hand, recent experiments~\cite{Madivala2009} show that capillary forces between elongated particles floating on spherical interfaces can have important consequences for stabilizing so-called Pickering emulsions, which are formed by particle-covered droplets (e.g.,\ oil) in a solvent (e.g.,\ water).

Whereas considerable theoretical progress has been made in understanding capillary interactions at flat interfaces~\cite{Kralchevsky2000,Oettel2005,Dominguez2008a}, basic issues such as the balance of forces acting on the interface and the influence of the incompressibility of the liquid enclosed by spherical interfaces have not yet been fully resolved. Curved interfaces pose the additional difficulty~\cite{Wurger2006,Dominguez2007} that in the presence of external forces the condition of mechanical equilibrium demands to either fix the center of mass of the droplet by an external body force or to pin the droplet surface, e.g., to a solid plate. The experimentally relevant issues of the boundary conditions at the plate and of their influence on the pair potential between capillary \textit{monopoles} have been studied previously~\cite{Guzowski2010,Guzowski2011}. Here, we study \textit{higher} capillary multipoles and derive the corresponding effective pair potentials for pointlike particles in terms of an expansion in spherical harmonics. By using a finite element method~\cite{Brakke} we compare these analytical results with those obtained by the numerical minimization of the surface free energy of drops in the presence of ellipsoidal particles.

\section{Perturbation theory}
We consider a spherical, full, and spatially fixed droplet of radius $R_0$ with colloidal particles trapped at its surface with surface tension $\gamma$. We model the effect of particles in terms of an external surface pressure field $\Pi(\Omega)$ (see below) parameterized by spherical coordinates $\Omega=(\theta,\phi)$ on the unit sphere. The equilibrium shape of the droplet subjected to the pressure field $\Pi$ follows from minimizing the corresponding free energy functional $\mathcal{F}[\{v(\Omega)\}]$ expressed in terms of the dimensionless radial displacement of the interface $v(\Omega)=(r(\Omega)-R_0)/R_0$. In the limit of small interfacial gradients $|\nabla_a v|\ll 1$ one has~\cite{Wurger2006,Guzowski2010}
 \begin{equation}
  \dfrac{1}{\gamma R_0^2}\mathcal{F}[\{v(\Omega)\}]= \int\! d\Omega\, \left[\frac{1}{2}(\nabla_a v)^2 - v^2 - \big(\pi(\Omega)+\mu\big)v\right],
  \label{free_en_O2}
 \end{equation}
where $\nabla_a:=\boldsymbol{e}_{\theta}\partial_{\theta}+\dfrac{\boldsymbol{e}_{\phi}}{\sin\theta}\partial_{\phi}$ is the dimensionless \textit{a}ngular gradient on the unit sphere~\cite{comment_epsilon}. The first two terms in Eq.\ (\ref{free_en_O2}) represent the surface free energy, the third term corresponds to the work done by the dimensionless external pressure $\pi(\Omega)=\Pi(\Omega)R_0/\gamma$ in displacing the interface and the fourth term serves to implement the volume conservation. The value of the Lagrange multiplier $-\mu$ follows from imposing the volume constraint 
 \begin{equation}
  \int\! d\Omega\, v = 0;
  \label{volume_constr2}
 \end{equation}
$\gamma\mu/R_0$ can be identified with the shift in the internal pressure of the droplet with respect to the Laplace pressure $2\gamma/R_0$ of an unperturbed, perfectly spherical droplet. The stationary condition $\delta \mathcal{F}/\delta v = 0$ leads to the Euler-Lagrange equation
 \begin{equation}
  -\big(\nabla_a^2+2\big)v(\Omega)=\pi(\Omega)+\mu.
  \label{helmholtz_pi}
 \end{equation}
We expand both the deformation $v(\Omega)$ and the pressure field $\pi(\Omega)$ in spherical harmonics $Y_{lm}(\Omega)$ so that Eq.\ (\ref{helmholtz_pi}) turns into an infinite set of algebraic equations~\cite{Morse1993}:
\begin{equation}
	[l(l+1)-2]v_{lm}=\pi_{lm}+\mu\delta_{l0},
	\label{helmholtz_multipole}
\end{equation}
where $l=0,1,\ldots$ and $m=-l,\ldots,l$, and the spherical multipoles are defined as $X_{lm}:=\int \!d\Omega'\, X(\Omega')Y_{lm}(\Omega')$ for $X(\Omega)=v(\Omega)$ or $X(\Omega)=\pi(\Omega)$. The volume constraint in Eq.\ (\ref{volume_constr2}) implies $v_{00}=0$ and thus $\mu = -\pi_{00}$, which means that the internal pressure shift counterbalances the external pressure. According to Eq.\ (\ref{helmholtz_multipole}) the $l=1$ components of the deformation $v$ are undefined. This is consistent with the fact that those components describe translations of the whole droplet without any change in shape which do not change the free energy. On the other hand all $l=1$ components of the external pressure $\pi$ must cancel, reflecting the condition of balance of forces acting on the droplet. Indeed, the multipoles $\pi_{1-1},\pi_{10},\pi_{11}$ are proportional to the Cartesian components $f_x,f_y,f_z$ of the total force $\boldsymbol{f}$ acting on the droplet. Hence the condition $\boldsymbol{f}=0$ is equivalent to $\pi_{1m}=0$ with $m=-1,0,1$. 


\section{Green's function and balance of forces}
As mentioned above we consider a pressure field in the form of the superposition of $N$ pointlike forces $f_i$ pointing into directions $\Omega_i$:
\begin{equation}
	\pi(\Omega)=\sum_{i=1}^{N}q_i\delta(\Omega,\Omega_i),
	\label{decomposition1}
\end{equation}
where $q_i:=f_i/(\gamma R_0)$ and $\delta(\Omega,\Omega_i)=\delta(\theta-\theta_i)\delta(\phi-\phi_i)/\sin\theta_i$ is the Dirac delta distribution expressed in terms of spherical coordinates. The pressure field as a whole must obey the condition of balance of forces which can be expressed as $\sum_i^{N}\pi_{i,1m}=0$ for $m=-1,0,1$. The corresponding interface deformation can be written as the superposition of single particle contributions: 
\begin{equation}
	v(\Omega)=\sum_i^{N}q_iG(\Omega,\Omega_i),
	\label{v1}
\end{equation}
where $G(\Omega,\Omega')$ is Green's function describing the response of the interface in direction $\Omega$ to a point-force applied in direction $\Omega'$ and which, according to the constraints of constant liquid volume and fixed center of mass obeys the equation~\cite{Morse1993}
\begin{equation}
	-(\nabla_a^2+2)G(\Omega,\Omega')=\sum_{l\geq 2}\sum_{m=-l}^l Y^*_{lm}(\Omega)Y_{lm}(\Omega'),
	\label{greens_equation}
\end{equation}
where the right hand side is a modified Dirac delta function $\hat{\delta}(\Omega-\Omega')$ with the $l=0$ and $l=1$ components projected out. From integrating both sides of Eq.\ (\ref{greens_equation}) over the unit sphere it follows that the $l=0$ component of $G$ vanishes, which reflects the condition of constant volume $\int\! d\Omega\, G(\Omega,\Omega')=0$. The $l=1$ component of $G$ also vanishes, which can be seen by multiplying both sides of Eq.\ (\ref{greens_equation}) by a radial vector $\boldsymbol{e}_r$, integrating over the unit sphere and using the fact that $\boldsymbol{e}_r$ can be expressed in terms of spherical harmonics with $l=1$ only. This reflects the condition that the center of mass of the droplet is fixed in space, which can be formulated as $\int\!d\Omega\,\boldsymbol{e}_r G(\Omega,\Omega')=0$. 
\begin{figure}
	\centering
	\psfragscanon
		\psfrag{p1}[l][l][1]{$\phi_1$}
		\psfrag{b}[c][c][1]{$\bar{\theta}$}
		\psfrag{p2}[l][l][1]{$\phi_2$}		
		\psfrag{o1}[r][l][1]{$\Omega_1=(0,0)$}
		\psfrag{o2}[c][c][1]{$\Omega_2$}	
	  	\psfrag{l}[c][c][1]{$\Lambda_{12}$}
	  	\psfrag{x}[c][c][1]{$x$}
		\psfrag{y}[c][c][1]{$y$}
		\psfrag{z}[c][c][1]{$z$}
	  	\psfrag{x1}[c][c][1]{$x'$}
		\psfrag{y1}[c][c][1]{$y'$}
		\psfrag{z1}[c][c][1]{$z,z'$}
	  	\psfrag{x2}[c][c][1]{$x''$}
		\psfrag{y2}[c][c][1]{$y''$}
		\psfrag{z2}[c][c][1]{$z''$}
		\includegraphics[width=0.45\textwidth]{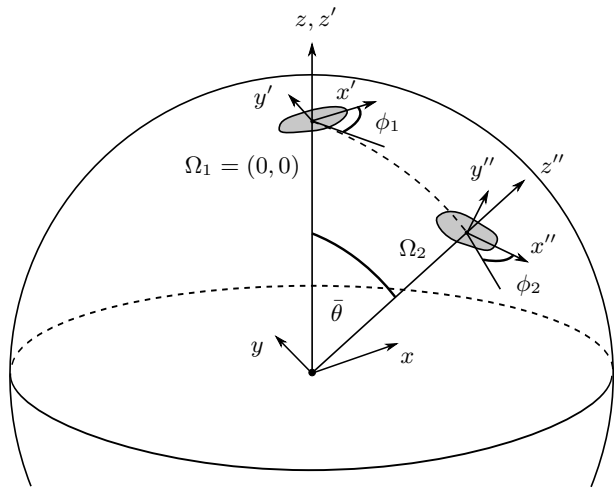}
	\psfragscanoff
		\caption{Angular configuration of the pressure distributions symbolically represented as gray patches centered at the directions $\Omega_1=(0,0)$ (north pole) and $\Omega_2$ on the unit sphere. The angles $\phi_1$ and $\phi_2$ represent orientations of the local coordinate frames $x'y'z'$ and $x''y''z''$ associated with the pressure distributions (see main text).}
	\label{sphere}
\end{figure}

According to the above considerations the pressure field due to a \textit{single} point-force violates the balance of forces acting on the droplet. Therefore, if one wants to cast the solution still into in the form given by Eq.\ (\ref{v1}) with $N=1$, there must be an additional pressure field counterbalancing the single point-force. Inserting  $v(\Omega)=G(\Omega,\Omega')$ into Eq.\ (\ref{helmholtz_pi}) and using Eq.\ (\ref{greens_equation}), one obtains the corresponding pressure field $\pi$ in the form
\begin{equation}
 \pi(\Omega) =\delta(\Omega,\Omega')+\pi_{CM}(\Omega,\Omega'),
\label{delta2}
\end{equation}
with the effective pressure $\pi_{CM}(\Omega,\Omega')=-(3/(4\pi))\cos\bar{\theta}$, where $\bar{\theta}$ is the angle between the directions $\Omega$ and $\Omega'$. The latter pressure must emerge in order to cancel the $l=1$ component of the external pressure $\pi_{ext}(\Omega)=\delta(\Omega,\Omega')$. Therefore $\pi_{CM}$ corresponds to a body force fixing the center of mass ($CM$) of the droplet without contributing to the deformation field $v$.

The expression in Eq.\ (\ref{v1}) for the deformation under the action of point-forces can be generalized to the case of a pressure field $\pi_{ext}(\Omega)$ due to an arbitrary continuous force distribution:
\begin{equation}
	v(\Omega)=\int\!d\Omega'\, \pi_{ext}(\Omega')G(\Omega,\Omega').
\label{upi}
\end{equation}
The total pressure $ \pi$ acting on the interface again has an additional component $\pi_{CM}$ accomplishing the fixing of the center of mass, i.e., $\pi=\pi_{ext}+\pi_{CM}$. One can express the free energy in terms of Green's function by integrating by parts Eq.\ (\ref{free_en_O2}), by using Eq.\ (\ref{helmholtz_pi}) and the divergence theorem, and finally by inserting $v(\Omega)$ as given by Eq.\ (\ref{upi})~\cite{Morse1993}:
\begin{equation}
	\dfrac{1}{\gamma R_0^2}F = -\dfrac{1}{2}\int\!d\Omega\, \int\!d\Omega'\, \pi_{ext}(\Omega)G(\Omega,\Omega')\pi_{ext}(\Omega').
\label{free_en_sph}
\end{equation} 

\section{Capillary interactions}
We study the pressure $\pi_{ext}$ localized around two different directions $\Omega_1$ and $\Omega_2$. To this end we introduce the following decomposition:
\begin{equation}
 \pi_{ext}(\Omega)=\pi_1(\hat{R}_1^{-1}\Omega)+\pi_2(\hat{R}_2^{-1}\Omega),
 \label{decomposition}
\end{equation}
where $\hat{R}_1$ ($\hat{R}_2$) denotes the rotation transforming the original coordinate frame $xyz$ into the coordinate frame $x'y'z'$ ($x''y''z''$), referred to as $O_1$ ($O_2$), associated with the pressure distribution $\pi_1$ ($\pi_2$) (see Fig.\ \ref{sphere}). We use the parameterization in terms of Euler angles, in which an arbitrary rotation $\hat{R}$ can be composed of a rotation around the $z$-axis by the angle $\alpha$, around the (rotated) $y$-axis by the angle $\beta$, and finally around the (rotated) $z$-axis by the angle $\gamma$. 
Under the rotation $\hat{R}(\alpha,\beta,\gamma)$ of the coordinate frame the coordinates transform according to $\hat{R}^{-1}$, as indicated in Eq.\ (\ref{decomposition}). 

The total free energy can be written as $F = F_{1,self} + F_{2,self} + \Delta F$, where $F_{i, self}= -[f_i^2/(4\pi\gamma)]\ln(R_0/a_i)+O(1)$ is the self-energy of particle $i$~\cite{Guzowski2010}, which does not depend on the relative position of the particles on the droplet but depends on the particle size $a_i$. For a particle located at the north pole this is defined as the smallest distance for which $\pi_i(\theta,\phi)=0$ for all $\theta>a_i/R_0$ and $\phi\in[0,2\pi)$; this corresponds to the solid angle circumscribing the particle-liquid interface. The interaction free energy $\Delta F$ is given by the cross-terms in Eq.\ (\ref{free_en_sph}) with $\pi$ from Eq.\ (\ref{decomposition}):
\begin{multline}
 \dfrac{1}{\gamma R_0^2}\Delta F = -\int \!d\Omega \int \!d\Omega'\, \pi_1(\hat{R}_1^{-1}\Omega)G(\Omega,\Omega')\times\\
\times\pi_2(\hat{R}_2^{-1}\Omega')
	= -\int \!d\Omega \int \!d\Omega'\, \sum_{l,m}\pi_{1,lm}Y^*_{lm}(\hat{R}_1^{-1}\Omega)\times\\
	\times\sum_{j,n} g_{j}Y^*_{jn}(\Omega)Y_{jn}(\Omega')\sum_{k,m'}\pi_{2,km'}Y^*_{km'}(\hat{R}_2^{-1}\Omega'),
\label{delta_f}
\end{multline}
where  
\begin{equation}
 g_l=\left\{
 \begin{array}{lll}
 0			&\text{for} &l=0,1\\
 \dfrac{1}{l(l+1)-2}	&\text{for} &l\geq 2
 \end{array} \right.
 \label{gl}
\end{equation}
are the spherical harmonics coefficients of $G(\Omega,\Omega')$. Without loss of generality we can take $\Omega_1=0$ such that the reference frame $O_1$ coincides with the original reference frame $xyz$ (see Fig.\ \ref{sphere}), which implies that $\hat{R}_1=\mathbbm{1}$. 

Spherical harmonics transform under the representation of the group of rotations according to $Y_{lm}(\hat{R}^{-1}\Omega)=\sum_{m'=-l}^{l}D^{\, l}_{m',m}(\hat{R})Y_{lm'}(\Omega)$, where $D^{\, l}_{m',m}$ is the Wigner $D$-matrix and reads (adopting the convention used in Ref.~\cite{Brink_book})
\begin{equation}
 D^{\, l}_{m',m}(\hat{R})=D^{\, l}_{m',m}(\alpha,\beta,\gamma)=e^{-im'\alpha}d^{\, l}_{m',m}(\beta)e^{-im\gamma},
\end{equation}
where $d^{\, l}_{m',m}(\beta)$ is known as the Wigner (small) $d$-matrix. We use the parameterization in terms of the orientations $\phi_1$ and $\phi_2$ of the coordinate frames $O_1$ and $O_2$, respectively, relative to the great circle connecting the points $\Omega_1=0$ and $\Omega_2$ on the unit sphere (see Fig.\ \ref{sphere}). The rotation $\hat{R}_2$ is then parameterized by the triad of the Euler angles $(2\pi-\phi_1,\bar{\theta},\phi_2)$. By using the orthogonality  $\int \!d\Omega Y_{lm}(\Omega)Y_{l'm'}^*(\Omega)=\delta_{ll'}\delta_{mm'}$ of the spherical harmonics, and due to the identities $Y_{lm}^*(\Omega)=(-1)^mY_{l,-m}(\Omega)$ and ${D^{\, l}_{m',m}}^*(\alpha,\beta,\gamma)=(-1)^{m-m'}D^{\, l}_{-m',-m}(\alpha,\beta,\gamma)$ we finally obtain
\begin{multline}
 \dfrac{1}{\gamma R_0^2}\Delta F = -\sum_{l\geq2} \sum_{m=-l}^l\sum_{m'=-l}^l\pi_{1,lm}\,\pi_{2,lm'}\,(-1)^{m'}\,g_l\times\\
\times 
d^{\, l}_{m,-m'}(\bar{\theta})\,e^{i(m\phi_1+m'\phi_2)}.
\label{multipole_expansion}
\end{multline}


\subsection{Limit of small particles}
Having in mind colloidal particles of radius $a$ as sources of the effective surface pressure $\pi$, we would like to relate the spherical multipoles $\pi_{lm}$, $l\geq0$, $m=-l,\ldots,l$ to prescribed forces, torques, and higher capillary multipoles defined for the case of a flat interface. 
If $a\ll R_0$ the interface can be treated as to be locally flat in the close neighborhood $\Delta\Omega$ of angular extent $O(a/R_0)$ around each particle separately. Assuming for reasons of simplicity that $\Delta\Omega$ is centered at the $z$-axis and approximating it by a circular disk $D(a)$ of radius $a$ in the tangent plane we can use the asymptotic form of spherical harmonics~\cite{Edmonds_book}
\begin{equation}
	Y_{lm}(\theta,\phi) \xrightarrow[\theta\rightarrow0]{}i^{|m|+m}A_{l|m|}\theta^{|m|} e^{im\phi},
\end{equation}
where
\begin{equation}
	A_{ln}=\sqrt{\dfrac{2l+1}{4\pi}\dfrac{(l+n)!}{(l-n)!}}\dfrac{1}{2^{n}n!},\quad n>0,
	\label{Alm}
\end{equation}
in order to obtain 
\begin{multline}
	\pi_{lm}=\int_{\Delta\Omega}\!d\Omega'\, \pi(\Omega') Y_{lm}(\Omega')\
	\xrightarrow[a/R_0\rightarrow0]{}\\
	i^{|m|+m}A_{l|m|}\left(\dfrac{a}{R_0}\right)^{|m|+1}Q_{|m|}e^{im\tilde{\phi}_{|m|}},
	\label{Qlm-Qm} 
\end{multline}
where we have introduced the dimensionless moduli
\begin{equation}
	Q_{n}:=\left|\int_{D(a)}\!\dfrac{d^2x'}{a^2}\,\dfrac{a\Pi(z')}{\gamma} \left(\dfrac{z'}{a}\right)^n \right|
	\label{Qnsph}
\end{equation}
and the phase
\begin{equation}
	\tilde{\phi}_{n}:=\dfrac{1}{n}\,\text{arg}\int_{D(a)}\!\dfrac{d^2x'}{a^2}\,\dfrac{a\Pi(z')}{\gamma} \left(\dfrac{z'}{a}\right)^n,
	\label{phin}
\end{equation}  
where $n\geq0$, $z'$ denotes a complex number, and $\Pi(\boldsymbol{x}')\equiv\gamma\pi(\Omega'(\boldsymbol{x}'))/R_0 $, with $\boldsymbol{x'}(\Omega')$ being a projection onto the plane tangent to the reference sphere at $\Omega_i$. From Eq.\ (\ref{Qnsph}) it follows that the moduli $Q_{n}$ are the usual capillary multipoles defined at a locally flat interface (see Refs.~\cite{Dominguez2008a,Dominguez_book}). Accordingly, due to the conditions of force and torque balance on a flat interface~\cite{Dominguez2008a}, $Q_{0}$ represents a ``capillary monopole'', i.e., the total external force acting on the particle, $Q_{1}$ a ``capillary dipole'', i.e., the total external torque. In the case of a free particle, $Q_{2}$ represents the lowest non-vanishing multipole. Furthermore, we note that for each particle $i$ one can always choose the orientation $\phi_i$ of the coordinate frame $O_i$ such that the phase of the order $n$ vanishes, i.e., $\tilde{\phi}_{i,n}=0$. We can write the interaction energy $\Delta F_{nn'}$ for a pair of capillary multipoles $Q_{n}$ and $Q'_{n'}$ of arbitrary orders $n\geq0$ and $n'\geq0$ as a sum of terms such that in Eq.\ (\ref{multipole_expansion}) the sum over $l$ is taken under the constraint $l\geq\max \{2,n,n'\}$ with $m$ and $m'$ fixed to $m=\pm n$ and $m'=\pm n'$; this implies $\Delta F = \sum_{n,n'=0}^{\infty}\Delta F_{nn'}$. Accordingly, the interaction free energy $\Delta F_{nn'}$ for a pair of capillary multipoles $Q_n$ and $Q_{n'}$ scales with the droplet radius $R_0$ as $\Delta F_{nn'}\sim \gamma a^{n+1}a'^{\,n'+1}/R_0^{n+n'}$. Note, that in the case of two monopoles ($n=0$ and $n'=0$) the interaction does not depend on $R_0$ but only on the moduli $Q_{0}$ and $Q'_{0}$.

The Wigner d-matrix in Eq.\ (\ref{multipole_expansion}) can be expressed in terms of the Jacobi polynomials $P_n^{(\sigma,\rho)}(\cos\bar{\theta})$~\cite{Edmonds_book}:
\begin{multline}
d^{\, l}_{m,m'}(\bar{\theta})=\left[\dfrac{(l+m')!(l-m')!}{(l+m)!(l-m)!}\right]^{1/2}
\left(\sin\dfrac{\bar{\theta}}{2}\right)^{m'-m}\times\\
\times
\left(\cos\dfrac{\bar{\theta}}{2}\right)^{m'+m}P^{(m'-m,m'+m)}_{l-m'}(\cos\bar{\theta}). 
\end{multline}
\begin{widetext}
By using certain properties of the Jacobi polynomials (see, e.g., Ref.~\cite{Edmonds_book}) we finally obtain
\begin{multline}
 	\dfrac{\Delta F_{nn'}}{\gamma a a'}\xrightarrow[a/R_0,a'/R_0\rightarrow0]{}
	\dfrac{Q_{n}Q'_{n'}}{(-2)^{n+n'+1}n!n'!\pi}\dfrac{a^{n}a'^{\,n'}}{R_0^{n+n'}}
	\sum_{l\geq \max\{2,n,n'\}}\dfrac{(2l+1)}{(l+2)(l-1)}\times\\ 
	\vspace*{-0.5cm}
	\times\left\{
\begin{array}{l}
		\dfrac{(l+n')!}{(l-n)!}\Bigg[(-1)^{n}\cos(n\phi_{1}+n'\phi_{2})\left(\cos\dfrac{\bar{\theta}}{2}\right)^{n'-n}\left(\sin\dfrac{\bar{\theta}}{2}\right)^{n'+n} P_{l-n'}^{(n'+n,n'-n)}(\cos\bar{\theta}) \\  
		\hspace*{-0.2cm}
		\begin{array}{lll}
		      \ + \cos(n\phi_{1}-n'\phi_{2})\left(\cos\dfrac{\bar{\theta}}{2}\right)^{n'+n}\left(\sin\dfrac{\bar{\theta}}{2}\right)^{n'-n}P_{l-n'}^{(n'-n,n'+n)}(\cos\bar{\theta})\Bigg],                                     & n>0, & n'>0,\\
		      (-1)^n\cos(n\phi_{1})P_l^n(\cos\bar{\theta}),          & n>0, & n'=0,\\
		      2^{-1}P_l(\cos\bar{\theta}),                                                                  & n=0, & n'=0,
		\end{array}
\end{array}\right.
	\label{Fn1n2}
\end{multline}
\noindent where $P_l$ and $P_l^n$ are the Legendre and the associated Legendre polynomials, respectively.
\end{widetext}

\subsection{Effective interaction potentials}
For a first discussion, we evaluate the interaction energy in Eq.\ (\ref{Fn1n2}) for $a=a'$ and $n'=n$. 
In the case of two identical pointlike \textit{monopoles}, $n=n'=0$, 
one obtains~\cite{Prudnikov_book2}
\begin{multline}
\frac{\Delta F_{00}(\bar{\theta})} {\gamma a^2}\xrightarrow[a/R_0\rightarrow0]{}\\
\frac{Q_0^2}{4\pi}\left[\frac{1}{2} + \frac{4}{3}\cos\bar{\theta} + 2\cos\bar{\theta}\ln\left(\sin\dfrac{\bar{\theta}}{2}\right) \right],
\label{greens_function}
\end{multline}
which gives back Green's function $G(\bar{\theta})\equiv-\Delta F_{00}(\bar{\theta})/(\gamma a^2 Q_0^2)$ as given in Ref.~\cite{Morse1993}. We note that according to Eq.\ (\ref{greens_function}) one has $G(\bar{\theta}\rightarrow0)\rightarrow -(1/(2\pi))\ln\bar{\theta}$, which renders the deformation $v(r)=-(1/(2\pi))\ln(r/R_0)$ due to a pointlike force, known for a flat interface, where $r=\bar{\theta}R_0\ll R_0$ is the arc length (see the dashed line in Fig.~\ref{sphere}) and $R_0$ plays the role of the capillary length. 

In the case of pointlike \textit{dipoles}, $n=n'=1$, one can evaluate the series involving the Jacobi polynomials by using their generating function (see the Appendix):
\begin{multline}
	\dfrac{1}{\gamma a^2}\Delta F_{11}(\bar{\theta},\phi_1,\phi_2) \xrightarrow[a/R_0\rightarrow0]{}\dfrac{Q_1^2}{8\pi}\left(\dfrac{a}{R_0}\right)^2\times \\
\times
\Big[ \cos(\phi_1+\phi_2)f_+(\bar{\theta}) + \cos(\phi_1-\phi_2)f_-(\bar{\theta}) \Big]
\label{DF11-final} 
\end{multline}
where $\phi_1$ and $\phi_2$ are the orientations of the particles as indicated in Fig.~\ref{sphere} and chosen such that the phases $\tilde{\phi}_{i,n}$ defined in Eq.\ (\ref{phin}) vanish for $n=1$, and where
\begin{align}
 f_+(\theta):=&\dfrac{1}{\sin^{2}(\theta/2)} - 4\sin^2\dfrac{\theta}{2}\ln\left(\sin\dfrac{\theta}{2}\right) - \dfrac{20}{3}\sin^2\dfrac{\theta}{2} + 2,\\
 f_-(\theta):=&4\left(\cos\dfrac{\theta}{2}\right)^2\ln\left(\sin\dfrac{\theta}{2}\right) + \dfrac{20}{3}\cos^2\dfrac{\theta}{2}.\label{f-}
\end{align}
We note that the dependences on the angular separation $\bar{\theta}$ and on the orientations $\phi_1,\phi_2$ do not factorize. Therefore, in order to minimize the free energy, the particles adopt orientations which depend on $\bar{\theta}$. According to Eq.\ (\ref{DF11-potential}) below one can distinguish three branches of the free energy (see the dotted green lines in Fig.~\ref{fig:potentials}) corresponding to three different minimal orientations, out of which the one with the lowest free energy is the equilibrium one. Assuming that the rotational relaxation of the particles is much faster than the translational one, an effective interaction potential can be obtained by minimizing the free energy with respect to $\phi_1$ and $\phi_2$:
\begin{multline}
	\dfrac{1}{\gamma a^2}\min_{\{\phi_1,\phi_2\}} \left\{\Delta F_{11}(\bar{\theta},\phi_1,\phi_2)\right\} \xrightarrow[a/R_0\rightarrow0]{}\\
\dfrac{Q_1^2}{8\pi}
\left\{\begin{array}{llll}\vspace*{0.1cm}
              -f_+(\bar{\theta})+f_-(\bar{\theta}), &\text{for}       & \bar{\theta}<\bar{\theta}_0,                   & \,\,\uparrow\,\,\,\uparrow \\ \vspace*{0.1cm}
              -f_+(\bar{\theta})-f_-(\bar{\theta}), &\text{for}       & \bar{\theta}_0<\bar{\theta}<\bar{\theta}_1, & \rightarrow\leftarrow \\ \vspace*{0.1cm}
              \quad f_+(\bar{\theta})-f_-(\bar{\theta}), &\text{for}  & \bar{\theta}>\bar{\theta}_1,                   & \,\,\uparrow\,\,\,\downarrow
             \end{array}\right.
\label{DF11-potential} 
\end{multline}
where $\bar{\theta}_0=\arccos[1-2\exp(-10/3)]\simeq0.38$ and $\bar{\theta}_1\simeq1.83$ are the zeros of $f_-(\bar{\theta}_0)=0$ and $f_+(\bar{\theta}_1)=0$, respectively; the arrows indicate the minimal orientational configuration (e.g., $\uparrow\,\,\,\uparrow$ corresponds to $(\phi_1=\pi/2,\phi_2=\pi/2)$ etc). 

In the case of pointlike \textit{quadrupoles}, $n=n'=2$, one obtains 
\begin{multline}
 	\dfrac{1}{\gamma a^2}\Delta F_{22}(\bar{\theta},\phi_1,\phi_2)  \xrightarrow[a/R_0\rightarrow0]{}\\ -\dfrac{3Q_2^2}{64\pi} \left(\dfrac{a}{R_0}\right)^4 \cos(2\phi_1+2\phi_2) \dfrac{1}{\sin^4(\bar{\theta}/2)}.
\label{DF22-final}
\end{multline}
Distinct from the case of dipoles, the free energy has a minimum for $\phi_1+\phi_2=k\pi,\, k=0,1,\ldots$ (so that $\cos(2\phi_1+2\phi_2)=1$), independent of $\bar{\theta}$. This corresponds to a monotonic attraction (see the solid blue line in Fig.~\ref{fig:potentials}). 

\section{Discussion}
\begin{figure}
	\centering
	\psfragscanon
		\psfrag{x}[l][l][1]{$\bar{\theta}$}
		\psfrag{x1}[l][l][1]{$\bar{\theta}_1$}
		\psfrag{x0}[l][l][1]{$\bar{\theta}_0$}
		\psfrag{as}[l][l][1]{$\bigg\}$ {\small flat}} 
		\psfrag{y}[c][c][1]{$(R_0/a)^{2n}\times\min_{\phi_1,\phi_2}\{\Delta F_{nn}\}/(\gamma a^2Q_n^2)$}
		\psfrag{n0}[l][l][1]{$n=0$}
		\psfrag{n1}[l][l][1]{$n=1$}
		\psfrag{n2}[l][l][1]{$n=2$}
		\begin{overpic}[width=0.45\textwidth]{fig02}
		 \put(65,8.5){\includegraphics[width=0.15\textwidth]{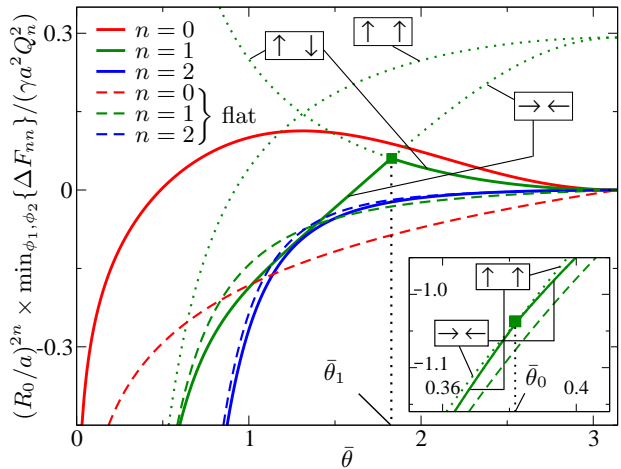}}
		\end{overpic}
	\psfragscanoff
		\caption{Rescaled effective interaction potentials (Eqs.\ (\ref{greens_function})-(\ref{DF22-final}), solid lines) for two capillary multipoles of the same order $n=0,1,2$, each shifted as a whole such that $\Delta F_{nn}(\bar{\theta}=\pi)=0$. The dashed lines correspond to the following expressions: $[\ln(\bar{\theta})-\ln(\pi)]/(2\pi)$ for $n=0$, $[1/\pi^2-1/\bar{\theta}^2]/(2\pi)$ for $n=1$, and $3[1/\pi^4-1/\bar{\theta}^4]/(4\pi)$ for $n=2$, representing for comparison the case of a flat interface with the equivalent spatial separation $d=R_0\bar{\theta}$ of the particles. As expected, for $\bar{\theta}\rightarrow0$ each full  curve approaches its corresponding dashed curve. The green dotted lines correspond to three metastable branches of the free energy for $n=1$ (see Eq.\ (\ref{DF11-potential})) with two relevant intersection points (at $\bar{\theta}_1\simeq1.83$, where $f_+(\bar{\theta}_1)=0$, and $\bar{\theta}_0\simeq0.38$, where $f_-(\bar{\theta}_0)=0$) indicated by filled squares. In each interval $[0,\bar{\theta}_0]$, $[\bar{\theta}_0,\bar{\theta}_1]$, and $[\bar{\theta}_1,\pi]$ the full green curve corresponds to the lowest of the three branches and as such describes thermal equilibrium.}
	\label{fig:potentials}
\end{figure}

We discuss the influence of the curvature of the interface on the interaction potentials by comparing the expressions for the curved interfaces (Eqs.\ (\ref{DF11-final})-(\ref{DF22-final}), solid lines in Fig.\ \ref{fig:potentials}) with the well established results for the case of a flat interface (see, e.g., Ref.~\cite{Dominguez_book}, dashed lines in Fig.\ \ref{fig:potentials}), with the corresponding spatial separation $d$ between the particles given by $d=R_0\bar{\theta}$. A striking difference is the non-monotonicity for curved interfaces of the free energy for monopole-monopole and dipole-dipole interactions, which in both cases leads to short-ranged attraction and long-ranged repulsion. These peculiarities can be traced back to the presence of an external body force necessary for fixing the center of mass. For $n=0,1$ the expected reduction of the results for curved interfaces to those for a flat interface occurs only for very small angular separations. This is not the case for quadrupoles ($n=2$), for which at all separations there is no qualitative discrepancy and only a minor quantitative difference from the flat limit.

\begin{figure}
	\centering
	\psfragscanon
		\psfrag{x}[l][l][1]{$\bar{\theta}$}
		\psfrag{y}[c][c][1]{$(R_0/a)^{4}\times\Delta F^{(2)}_{ell}/(\gamma a^2Q_2^2)$}
		\psfrag{r5}[l][l][1]{$a/R_0=0.599$}
		\psfrag{r6}[l][l][1]{$a/R_0=0.5$}
		\psfrag{r8}[l][l][1]{$a/R_0=0.375$}
		\psfrag{2a}[l][l][0.9]{$2a$}
		\psfrag{2b}[l][l][0.9]{$2b$}
		\psfrag{ps}[l][l][0.9]{$\psi_i$}
		\psfrag{om}[l][l][0.9]{$\Omega_i$}
		\psfrag{h}[l][l][0.9]{$h_i$}
		\psfrag{r0}[c][c][0.9]{$R_0$}
		\psfrag{tp}[l][l][0.9]{$\theta_p$}
		\psfrag{2r}[c][c][1]{$2R_0$}
		\psfrag{blank}[l][l][1]{}
		\psfrag{q}[l][l][1]{point-quadrupoles}
		\psfrag{ss}[l][l][1]{$\Bigg\}$}
		\psfrag{tt}[l][l][1]{$\Big\}$}
		\begin{overpic}[width=0.45\textwidth]{fig04}
		  \put(53,48){\includegraphics[width=0.142\textwidth]{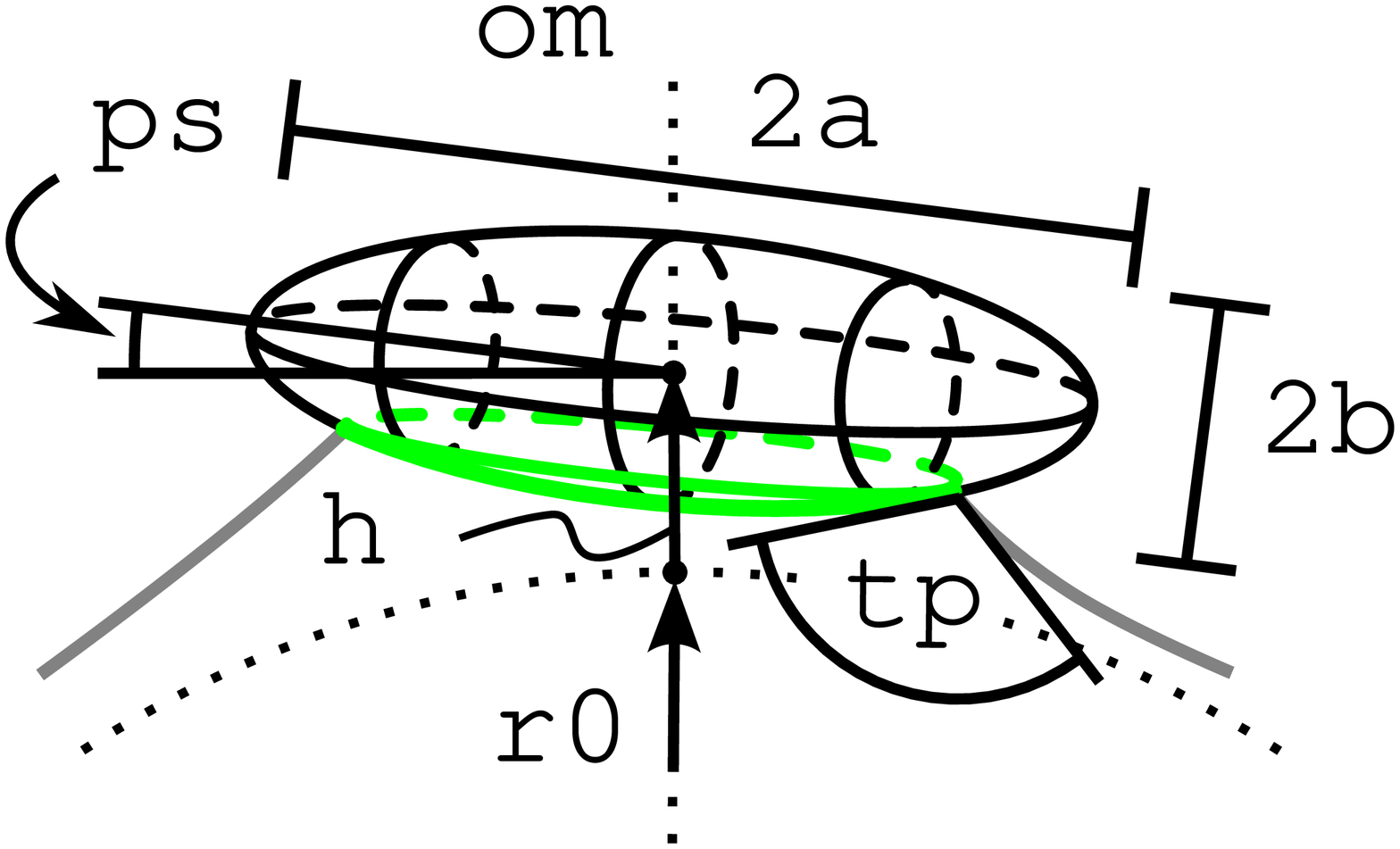}}
		  \put(70,33){\includegraphics[width=0.0675\textwidth]{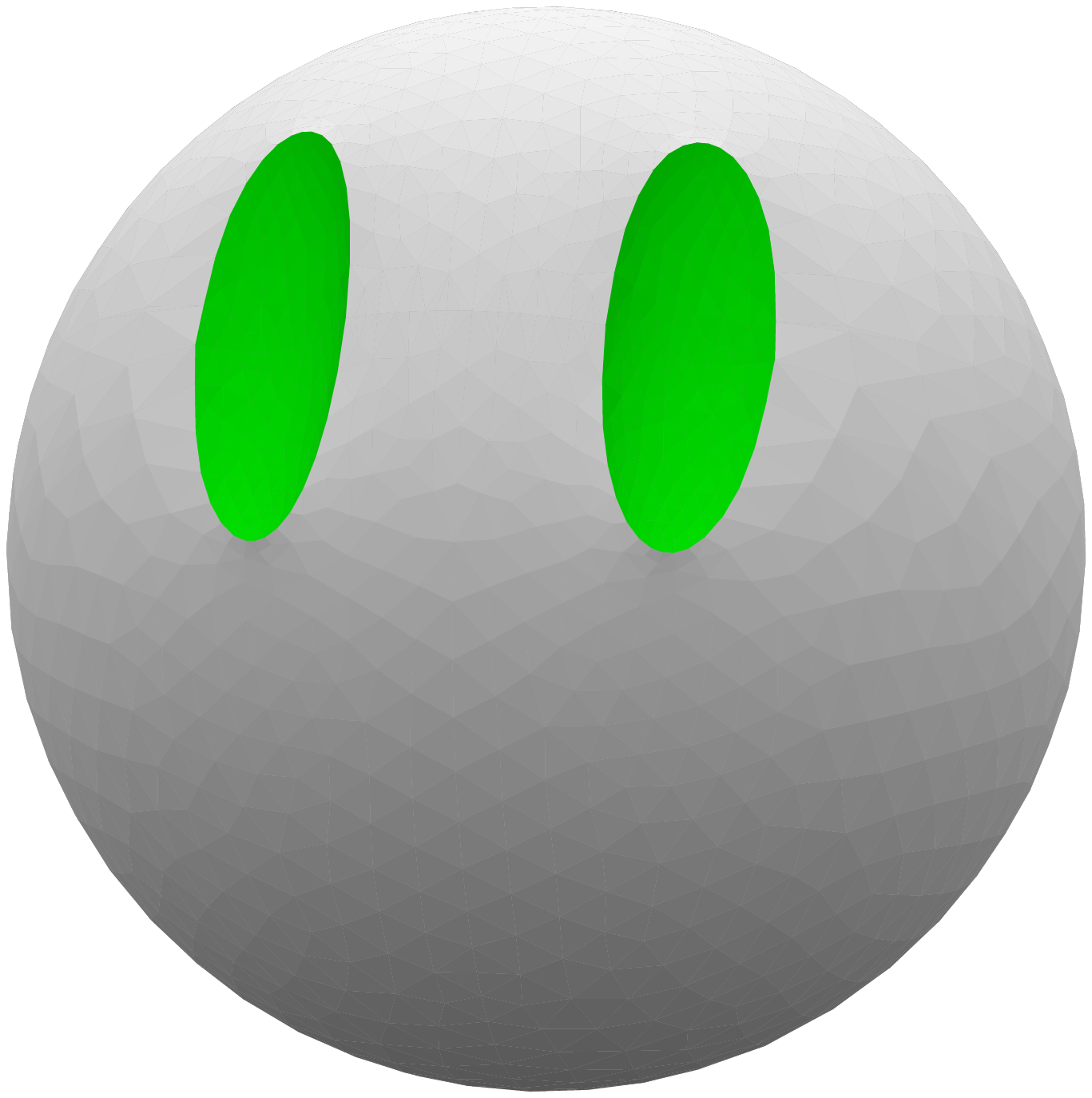}}
		  \put(70,15){\includegraphics[width=0.0675\textwidth]{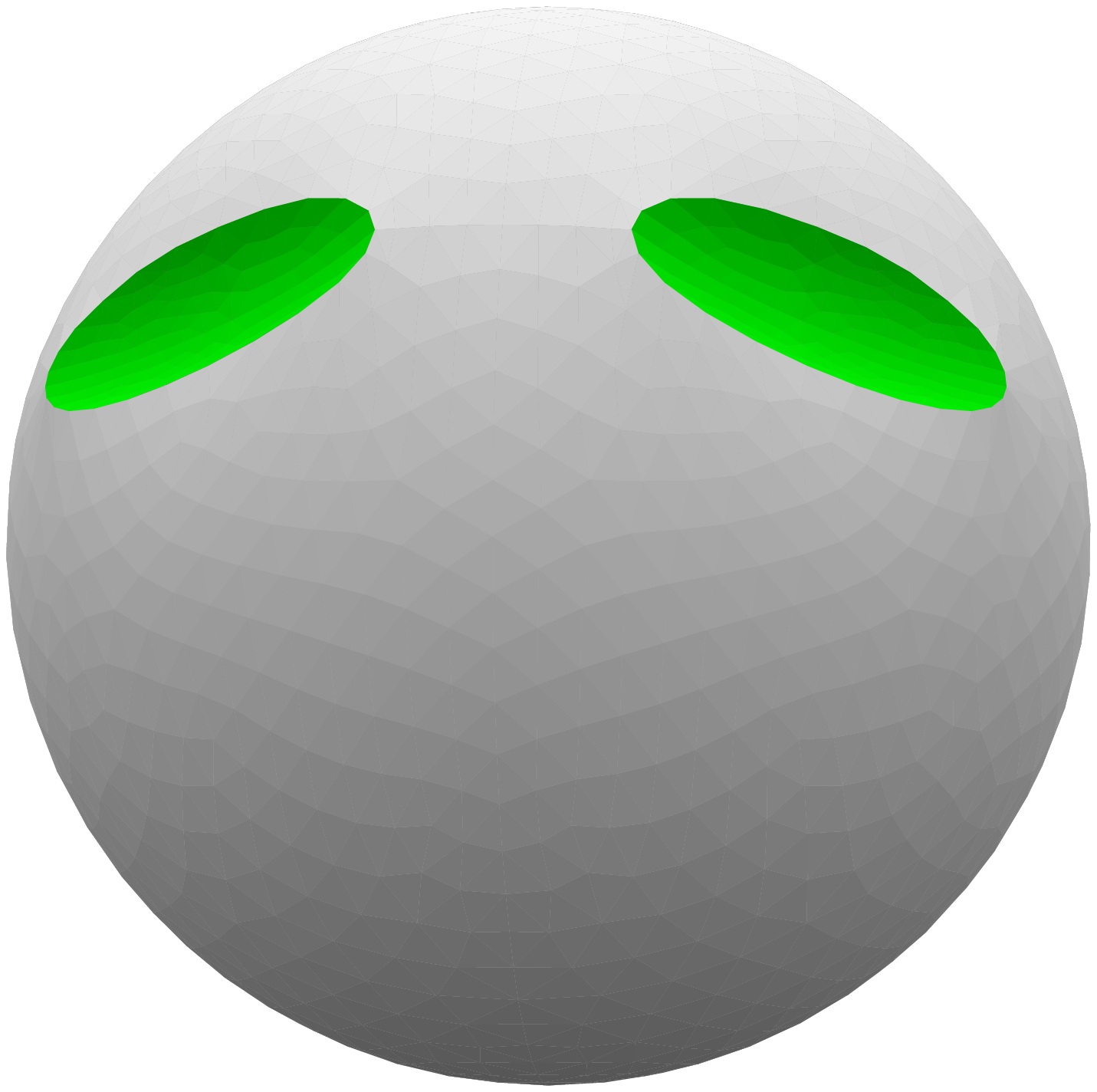}}
		  \put(86,33){\includegraphics[width=0.0336\textwidth]{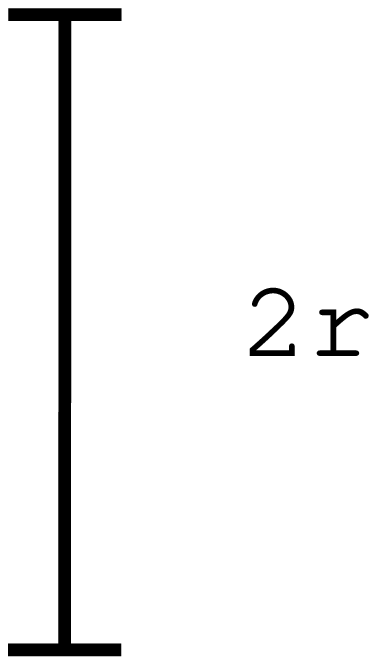}}
		\end{overpic}
	\psfragscanoff
		\caption{Rescaled effective interaction energy $\Delta F^{(2)}_{ell}$ for two ellipsoidal particles of aspect ratio $a/b=3$, contact angle $\theta_p=2\pi/3$, various droplet radii $R_0$, and two different orientational configurations: tip-to-tip and side-to-side (as visualized by the snapshots from the finite-element free energy minimization). In view of large relative numerical errors for small values of $\Delta F^{(2)}_{ell}$ the zero of the free energy is taken as the mean value of the data points with $\bar{\theta}>1.5$ (the error bars are comparable or smaller than the size of the symbols). The thick solid line corresponds to Eq.\ (\ref{DF22-final}) with $\phi_1=\phi_2=0$ or $\phi_1=\phi_2=\pi/2$ (in either case Eq.\ (\ref{DF22-final}) yields the same free energy). The capillary quadrupole of each of the particles is approximated by the expression $Q_2=2\pi\Delta u/a$; this relation is valid for a flat interface~\cite{Dominguez2008a} at large interparticle separations, with the undulation $\Delta u$ of the interface taken to be the one at the surface of a single ellipsoidal particle. As an approximation for a droplet we adopt the same relation, $Q_2(R_0) = 2\pi \Delta u(R_0)/a$, with the undulation $\Delta u\equiv\max_{\phi}r(\theta=a/R_0,\phi)-\min_{\phi}r(\theta=a/R_0,\phi)$ of the interface at the bottom part of a particle centered at the north pole of the droplet (see the green part of the upper inset). This undulation has been calculated numerically for three droplet radii: $\Delta u(a/R_0=0.599)/a=0.028$, $\Delta u(a/R_0=0.5)/a=0.031$, $\Delta u(a/R_0=0.375)/a=0.034$. $R_0$ is the radius of the droplet \textit{without} the particles. The immersion $h_i$ and the tilt angle $\psi_i$ of the long axis of each particle $i$ with respect to the plane perpendicular to the direction $\Omega_i$ are subjected to minimization. The colored lines end at small values of $\bar{\theta}$ because the particle surfaces touch each other for nonzero $\bar{\theta}$.} 
	\label{fig:numerical}
\end{figure}

Next, we compare our approximate analytic theory with the numerical minimization of the free energy for floating ellipsoidal particles with semi-axes $a$ and $b$, which we performed by using a finite element method~\cite{Brakke}. The numerical procedure consists of minimizing the free energy functional
\begin{multline}
	\mathcal{F}[\{\boldsymbol{r}(\Omega)\},h_1,h_2,\psi_1,\psi_2;\bar{\theta},\phi_1,\phi_2,\theta_p,a,b,V_l,\lambda]=\\
	= \gamma S_{lg} -\gamma\sum_{i=1,2}\cos\theta_{p,i}S_{pl,i} - \lambda(V-V_l),
	\label{functional}
\end{multline}
with respect to the shape of the interface $\{\boldsymbol{r}(\Omega)\}$, immersions of the particles $h_1,h_2$, and their tilts $\psi_1,\psi_2$ (relative to planes perpendicular to the directions $\Omega_1$ and $\Omega_2$, respectively, see the inset in Fig.\ \ref{fig:numerical}), while the angular separation $\bar{\theta}$ and the tangential orientations $\phi_1,\phi_2$ are kept fixed; $S_{lg}$ is the area of the liquid-gas interface, and $S_{pl,i}$ and $\theta_{p,i}$ are the area of the particle-liquid interface and the contact angle, respectively, of particle $i$. The last term in Eq.\ (\ref{functional}) ensures conservation of the liquid volume $V_l$ with $\lambda$ as the corresponding Lagrange multiplier. 

The effective potentials (Fig.~\ref{fig:numerical}) obtained numerically for prolate spheroids (i.e., elongated spheres) with aspect ratio $a/b=3$ and droplet radii $a/R_0=0.599,0.5,0.375$ differ for particles oriented tip-to-tip ($\phi_1=\phi_2=\pi/2$) and side-to-side ($\phi_1=\phi_2=0$). This signals the importance of capillary multipoles higher than only a quadrupole, because the point-quadrupole approximation yields the same free energy for both orientations (see Eq.\ (\ref{DF22-final})). This discrepancy is not surprising because the longer particle axes (which set the length scale $a$) are actually not very small compared with the droplet radius $R_0$ and thus in this case the point-particle limit is a rather crude approximation. Our numerical analysis is restricted to such small droplets due to the limited effectiveness of the finite-element method for larger drops caused by too large numerical errors. For the same reason we consider only particles with a large aspect ratio ($a/b=3$) for which the deformation of the interface is sufficiently large so that the numerical noise does not smear out the angular dependence of the free energy. Nonetheless, even in this regime our analytic theory captures the main qualitative features of the effective interaction potential between floating ellipsoids. 
Presumably the reason for this is that even for large particles the droplet shape deviates only slightly from a sphere. (Due to a fixed volume of the liquid  larger particles cause a larger displaced liquid volume $V_{displ}$; primarily this leads to an increased effective radius $\sim\sqrt[3]{V_l+V_{displ}}$ of the droplet and thus generates only a constant in the deformation field $v$.) Accordingly, away from the particles our perturbation theory remains applicable. Thus even for large particles the deformation of the droplet can be approximated by the superposition of contributions from pointlike quadrupoles substituting  the actual particles. 
The additional dependence on the orientations revealed by our numerical calculations, which goes beyond the quadrupole approximation, resembles the well established case of a flat interface, which has been analyzed experimentally~\cite{Loudet2005} and theoretically~\cite{Lehle2008} in terms of a multipole expansion. Those studies showed that the point-quadrupole approximation, within which tip-to-tip and side-to-side configurations have the same free energy, is reliable only either for particles with a very small eccentricity ($b\simeq a$) or at large spatial separations $d\ll a$. Otherwise, one should rather consider an expansion in terms of elliptic coordinates, from which it follows that the tip-to-tip configuration is energetically more favourable. This prediction agrees with the experimental observations~\cite{Loudet2005}. Our calculations indicate that the same configuration is the preferred one also in the case of spherical interfaces. However, we have not analyzed all the possible orientations. It is noteworthy that optimal orientations other then tip-to-tip and side-to-side have been observed at a flat interface for a pair of ellipsoidal particles of \textit{different} sizes ~\cite{Loudet2009}.  

We can also discuss the effective capillary potential $F^{(1)}_{ell}(\alpha,\phi_1;\theta_0)$ for a single capillary quadrupole on a \textit{sessile}, spherical caplike droplet forming a contact angle $\theta_0$ with the supporting substrate, where $\alpha$ is the angular position of the particle ($\alpha<\theta_0$) measured from the droplet symmetry axis and relative to the center of the sphere circumscribing the droplet, and $\phi_1$ the orientation of the particle. As before the tilt angle $\psi_1$ is taken to be the optimal one. In the case $\theta_0=\pi/2$ the method of images can be applied, analogous to the case of monopoles~\cite{Guzowski2010}. The substrate acts like a mirror and the free energy of a \textit{single} particle equals one half of the energy of \textit{two} particles on a \textit{full} droplet, i.e., 
\begin{multline}
\Delta F^{(1)}_{ell}(\alpha,\phi_1;\theta_0=\pi/2):=F^{(1)}_{ell}(\alpha,\phi_1;\theta_0=\pi/2)\\
-F^{(1)}_{ell}(0,\phi_1;\theta_0=\pi/2)=\\
=\pm\dfrac{1}{2}\Delta F_{22}(\pi-2\alpha,\phi_1,\pi-\phi_1),
\label{eq:DFell}
\end{multline}
where the $+$ sign corresponds to a free contact line at the substrate and the $-$ sign to a pinned one. (Since the particle is force-free, there are no additional images, distinct from the case of monopoles, see Ref.~\cite{Guzowski2010}.) For a free contact line Eq.\ (\ref{eq:DFell}) holds in general, i.e., for arbitrary particles (not necessarily ellipsoidal). In such a case $\Delta F_{22}$ is replaced by the actual interaction energy $\Delta F^{(2)}$ between the particles (possibly containing also higher multipoles). 
On the other hand, for a pinned contact line Eq.\ (\ref{eq:DFell}) holds only for pointlike particles in conjunction with Eq.\ (\ref{DF22-final}).

For a pointlike quadrupole, according to the expression for $\Delta F_{22}$ in Eq.\ (\ref{DF22-final}), the dependence of $\Delta F^{(1)}_{ell}$ on the orientation $\phi_1$ drops out so that within this approximation the particle is always attracted to a free contact line and repelled from a pinned one. Thus the effective potential is given by the black line in Fig.\ \ref{fig:numerical} with $\bar{\theta}=\pi-2\alpha$ (with an additional minus sign in the case of a pinned contact line). 

In the case of an actual ellipsoidal particle the free energy is no longer given by the rhs of Eq.\ (\ref{DF22-final}) so that the dependence on $\phi_1$ does no longer drop out but becomes important in the vicinity of the contact line at the substrate. As stated above, if this contact line is \textit{free} (still assuming $\theta_0=\pi/2$), due to symmetry arguments the free energy of a single particle on a sessile drop equals one half of the energy for two particles on a full droplet (as predicted by the method of images for the more restrictive case of pointlike quadrupoles), for which the full numerical results are plotted in Fig.\ \ref{fig:numerical}. Accordingly, for a fixed angular position $\alpha=\pi/2-\bar{\theta}/2$ of the particle its orientation perpendicular to the contact line (solid lines in Fig.\ \ref{fig:numerical}) is energetically more favorable. However, the global minimum of the free energy corresponds to the particle touching the substrate (i.e., $\alpha\rightarrow\pi/2$ so that $\bar{\theta}\rightarrow0$) and being oriented parallel to the contact line (dashed lines in Fig.\ \ref{fig:numerical} for $\bar{\theta}<0.5$). However, we cannot judge if the latter configuration remains the optimal one for arbitrary aspect ratios and contact angles at the particle. (This and analyzing the influence of the contact angle at the substrate actually call for a separate numerical study.) 

If the contact line at the substrate is \textit{pinned}, the global free energy minimum corresponds to the particle being positioned at the drop apex. In such a case, due to a large distance from the contact line with the substrate, the pointlike quadrupole approximation remains reliable. One can estimate the depth of this effective confining potential as follows. According to Eqs.\ (\ref{eq:DFell}) and (\ref{DF22-final}) with $Q_2=2\pi\Delta u/a$ and assuming $\Delta u/a\simeq 0.05$ (which  applies to an ellipsoidal particle of aspect ratio $a/b=3$ and contact angle $\theta_p=2\pi/3$, see the caption of Fig.\ \ref{fig:numerical}), the displacement of the particle, e.g., from $\alpha=0$ to $\alpha=\pi/4$, leads to a free energy increase of approximately $k_BT\times(a/\mu m)^2$ for a droplet as large as $R_0/a\simeq 10$ (assuming a surface tension $\gamma\simeq0.07 N/m$ which corresponds to a water droplet in air). Hence, one could experimentally measure the capillary confining potential, e.g., by observing the Brownian motion of micron-sized particles on sessile drops. Finally, we emphasize that the effects of confinement should be also taken into account in determining the effective \textit{pair} potentials between particles on \textit{sessile} drops (for the case of monopoles see Ref.~\cite{Guzowski2011}). In particular, the confinement leads to closer packing of particles on a droplet, which should not be mistaken for an additional attractive pair-interaction.

\section{Acknowledgements}
On of the authors (J.G.) acknowledges financial support from the Polish Ministry of Science (2010/2011) through grant nr IP2010 012270.

\section{Appendix: Calculation of series containing Jacobi polynomials}

In Eq.\ (\ref{Fn1n2}) for $n=n'=1$ the following series appears:
\begin{equation}
 S_1^{(\sigma_1,\rho_1)}(x):=
 \sum_{j\geq1}\dfrac{(j+1)(j+2)(2j+3)}{j(j+3)}P_j^{(\sigma_1,\rho_1)}(x)
\end{equation}
where $(\sigma_1,\rho_1)=(2,0)$ or $(0,2)$, and for $n=n'=2$: 
\begin{equation}
  S_2^{(\sigma_2,\rho_2)}(x):=
  \sum_{j\geq0}(j+2)(j+3)(2j+5)P_j^{(\sigma_2,\rho_2)}(x)
\end{equation}
where $(\sigma_2,\rho_2)=(4,0)$ or $(0,4)$. Those series can be evaluated using the generating
function~\cite{Abramowitz_book}:
\begin{multline}
  g^{(\sigma,\rho)}(x,z)=\dfrac{1}{R(1-z+R)^{\sigma}(1+z+R)^{\rho}}\\
 =\sum_{n=0}^{\infty}\dfrac{z^nP_n^{(\sigma,\rho)}(x)}{2^{\sigma+\rho}},
\end{multline}
where $R=\sqrt{1-2xz+z^2}$ and $|z|<1$. One finds the relationships
\begin{multline}
  S_1^{(\sigma_1,\rho_1)}(x)=2\int_0^1\!dy\, y^2 \int_0^y\!dz\,\times \\
\times z^{-2}\dfrac{\partial}{\partial z}z^{1/2}\dfrac{\partial}{\partial z}z^{2}\dfrac{\partial}{\partial z}z\left[2^{\sigma_1+\rho_1}g^{(\sigma_1,\rho_1)}(x,z)-1\right]
\end{multline}
and
\begin{multline}
 S_2^{(\sigma_2,\rho_2)}(x)=\\
 =2^{\sigma_2+\rho_2+1}\lim_{z\rightarrow1}\dfrac{\partial}{\partial z}z^{1/2}\dfrac{\partial}{\partial z}z^{2}
 \dfrac{\partial}{\partial z}z^2g^{(\sigma_2,\rho_2)}(x,z), 
\end{multline}
which can be evaluated with the results
\begin{align}
 &S_1^{(2,0)}(x)=\dfrac{4}{(1-x)^2}+\dfrac{4}{1-x}-2\ln\left(\dfrac{1-x}{2}\right)-\dfrac{20}{3},\\
 &S_1^{(0,2)}(x)=-2\ln\left(\dfrac{1-x}{2}\right)-\dfrac{20}{3},\\
 &S_2^{(4,0)}(x)=\dfrac{96}{(1-x)^4} \label{sum0},
\end{align}
and
\begin{equation}
 S_2^{(0,4)}(x)=0.
\end{equation}
Taking $x=\cos\bar{\theta}$ and inserting these results into Eq.\ (\ref{Fn1n2}) finally yields Eqs.\ (\ref{DF11-final})-(\ref{DF22-final}).


\begin{thebibliography}{99}
\bibitem[{com()}]{comment_epsilon}
\bibinfo{note}{Different from Eqs.\ (20)-(25) in Ref.~\cite{Guzowski2010} here we do not include the small parameter $\epsilon$ - introduced therein and associated with an external force - into the definition of the dimensionless quantities.}
\end{thebibliography}

\end{document}